\documentclass[structabstract]{aa}
\usepackage{graphicx}
\usepackage{txfonts}
\usepackage{url}
\begin{document}
   \title{Simulating flaring events in complex active regions driven by observed magnetograms}


   \author{M. Dimitropoulou,
          \inst{1}
          H. Isliker,
          \inst{2}
          L. Vlahos,
          \inst{2}
          \and
          M. K. Georgoulis
          \inst{3}
          }
   \institute{University of Athens, Department of Physics, GR-15483, Athens, Greece\\
              \email{michaila.dimitropoulou@nsn.com}
   \and
              University of Thessaloniki,Department of Physics, GR-54006, Thessaloniki, Greece\\
   \and
              Research Center of Astronomy and Applied Mathematics, Academy of Athens, GR-11527, Greece\\
             }




  \abstract
   {We interpret solar flares as events originating from active regions
   that have reached the Self Organized Critical state, by using a refined
   Cellular Automaton model with initial conditions derived from observations.
    }
   {We investigate whether the system, with its imposed physical elements,
   reaches a Self Organized Critical state and whether well-known statistical properties of flares,
   such as scaling laws observed in the distribution functions of characteristic parameters,
   are reproduced after this state has been reached.}
   {To investigate whether the distribution functions of total energy,
    peak energy and event duration follow the expected scaling laws, we first
    apply a nonlinear force-free extrapolation which reconstructs the three-dimensional
    magnetic fields from two-dimensional vector magnetograms.
    We then locate magnetic discontinuities exceeding a threshold in the Laplacian of the magnetic field. These
    discontinuities are relaxed in local diffusion events, implemented in the form of Cellular Automaton evolution rules.
    Subsequent loading and relaxation steps lead the system to Self Organized Criticality,
    after which the statistical properties of the simulated events are examined. Physical requirements, such as the
    divergence-free condition for the magnetic field vector, are approximately imposed on all elements of the model.}
   {Our results show that Self Organized Criticality is indeed reached when applying specific
    loading and relaxation rules. Power law indices obtained from the distribution
    functions of the modeled flaring events are in good agreement
    with observations. Single power laws (peak and total flare energy) as well as power laws with exponential cutoff and double power laws (flare duration) are obtained. The results are also compared with observational
    X-ray data from GOES satellite for our active-region sample.}
   {We conclude that well-known statistical properties of flares are reproduced after the system has reached
    Self Organized Criticality. A significant enhancement of our refined Cellular Automaton model
    is that it commences
    the simulation from observed vector magnetograms, thus facilitating energy calculation in physical units.
    The model described in this study remains consistent with fundamental physical requirements, and imposes
    physically meaningful driving and redistribution rules.}

   \keywords{Sun: corona --
                Sun: flares --
                Sun: photosphere
               }
\titlerunning{Simulating flares in active regions driven by observed magnetograms}
\authorrunning{Dimitropoulou et al.}
   \maketitle
%

\section{Introduction}

Solar flares are transient energy release events above solar Active Regions (ARs).
Populations of flares are known to exhibit robust statistical properties,
which have been repeatedly identified in numerous observations. In particular, specific flare
parameters have been consistently found to follow robust power laws with indices lying
in well-defined ranges. More specifically, a series of flare observations
(\cite{dat74,lin84,stu84,den85,vil87, cro93, bie94, bro95, pol02}) report that
the distribution functions of peak flux, total energy and event duration exhibit well-formed
scaling laws with exponents in the ranges of $(-1.59, -1.80)$, $(-1.39, -1.50)$ and
$(-2.25, -2.80)$ respectively.

This consistency of the power law indices identified in numerous independent studies
stimulated a new phenomenological approach in reproducing and modeling the statistical
behaviour of flaring activity. \cite{luh91} and \cite{luh93} were the first to construct
a simple model of solar flare occurrence, based on the assumption that the solar corona is in
a statistically stable Self Organized Critical (SOC) state. In this context, ARs are perceived as
nonlinear dissipative dynamical systems, externally driven by the photospheric velocity field.
Due to random shuffling of the coronal loops' footpoints in the photosphere,
localized instabilities are generated, which are responsible for the fragmented energy release in the solar corona.
The magnetic energy release simulated via Cellular Automaton (CA) modeling led to avalanche-like events.
This model allows instabilities, simulating current sheet disruption and Ohmic dissipation, when a
certain current density threshold is exceeded. An enhancement of the original SOC concept with respect to
the instability criteria and the corresponding relaxation was introduced by \cite{vla95} and \cite{geo95}.
Both suggested that the initial instability may trigger secondary ones, thus affecting sites beyond the closest
vicinity of the original event. Non-local treatment between flaring elements was also attempted by \cite{mac97}. In addition, \cite{geo96} constructed a refined
Statistical Flare Model, including both isotropic and anisotropic relaxation mechanisms as well as extended instability
criteria. This kind of modeling produced a double power-law scaling behaviour: the flatter power law resembled intermediate
and large flares, whereas the steeper one described low-energy events. An additional enhancement of this model lay in
the external driver simulation. In the mentioned study, the driver of the system followed a power law itself, thus mimicking the
instabilities triggered by the emerging magnetic flux from the convection zone in addition to photospheric shuffling.
Further extensions were introduced by \cite{geo98}, who presented a systematic study of the power law
indices' variability as a function of the driver's properties. In this refined Statistic Flare Model, Georgoulis \& Vlahos
attempted to model the stresses which are built-up randomly within ARs through a highly variable, inhomogenous external driver.
Although clearly deviating from the initial SOC models, the robust scaling laws in the flares' distribution functions
survived. Isliker et al. (\cite{isl00,isl01}) tried for the first time to associate the classical CA models' components with physical variables.
Magnetohydrodynamic (MHD) and CA approaches were connected through the physical interpretation of numerous CA elements, like the grid-variable,
the time step, the spatial discreteness, the energy release process and the role of diffusivity. This study revealed several
inconsistencies of the CA modeling, such as the uncontrolled value of the magnetic field
divergence ($\vec{\nabla\cdot\vec{B}}$) and the non-availability of secondary variables, such as the current density and the electric field.
Such weaknesses were treated by the Extended CA model (X-CA) introduced by Isliker et al. \cite{isl02}.

In this study we present a model which adopts the Lu \& Hamilton (1991) approach as starting point and to which several enhancements are made towards a more physical CA model that integrates various aspects of observed ARs and flares: first and foremost, the initial boundary and initial
conditions stem from observed vector magnetograms. This allows us to perform calculations in physical units, in direct comparison with observations (see for example the respective restrictions presented in \cite{geo01}). Time remains the only quantity expressed in arbitrary model units, as the photospheric vector magnetogram does not change during the simulation. An additional feature is that during the whole process (initial loading, relaxation
of magnetic discontinuities and further driving) the requirement $\vec{\nabla\cdot\vec{B}}\simeq{0}$ is explicitly imposed. For this purpose we have used a nonlinear force-free extrapolation method to generate the initial
conditions from observed magnetograms and impose instability criteria related
to actual physical processes. The magnetic field relaxation in the CA model follows the \cite{luh91}
principles. The driving process is also designed to obey specific rules which
do not violate known physical processes in the corona. The structure of this work is as follows. Section 2 describes the data used in this
study along with the necessary corrections imposed on them. Section 3 explains in detail all the modules comprising our model: first the extrapolation technique (EXTRA) along with the discontinuities' identification (DISCO) modules. Furthermore, the magnetic field relaxation module (RELAX) and finally the
driving module (LOAD) is presented, which may trigger further instabilities in the simulated AR, following rules which mimic specific physical processes. Section 4 presents our results and discusses our findings. Finally, Sect. 5 summarizes our conclusions.


\section{Data-set}

   Nonlinear force-free extrapolation techniques require vector
   magnetograms that are not as widely available as conventional line-of-sight
   magnetograms. Here we have created a database of 11 different AR
   vector magnetograms from the University of Hawaii Imaging Vector Magnetograph (IVM).

   IVM obtains Stokes images in photospheric lines with $7 pm$ spectral resolution, $1.1 arcsec$ spatial resolution
   ($\sim{0.55}$ arcsec per pixel) over a field of $4.7 arcmin^{2}$ and polarimetric precision of $0.1\%$ (\cite{mic96}).
   We use both fully-inverted and quick-look IVM data. Quick-look data have been obtained from the IVM Survey Data archive (available online at
   \url{http://www.cora.nwra.com/ivm/IVM-SurveyData}). The quick-look data reduction differs from the complete inversion in that it uses a
   simplified flat-fielding approach, takes no account of scattered or parasitic light, and no correction is
   attempted for seeing variations that occur during the data acquisition.


   In this study we use 1 fully inverted and 10 quick-look IVM vector magnetograms.
   To remove the intrinsic azimuthal ambiguity of $180^{o}$ we use the Non-Potential magnetic Field Calculation (NPFC) method
   of Georgoulis (\cite{geo05}). For computational convenience we further rebin the disambiguated magnetograms into a $32\times{32}$ regular grid.



\section{The model}

Our model consists of 4 separate modules. First we apply
the Wiegelmann (2008) optimization algorithm to our vector magnetograms in order to nonlinearly extrapolate the magnetic field
from the photospheric boundary (module ``EXTRA"). We thus construct a three-dimensional (3d) $32\times 32\times 32$ cube, within which the
magnetic field is determined.
Second, we identify the sites within our cubic grid that exceed a threshold in the magnetic field Laplacian (module ``DISCO").
If unstable sites are found, we force the vicinity of the unstable location to undergo a magnetic-field restructuring. This redistribution is governed by specific rules, which do not violate basic physical laws.
Under suitable conditions, the onset and relaxation of an initial instability may trigger a cascade of similar events in an avalanche-type manner. It is clear, therefore, that the wider vicinity, up to the entire system, may participate in this process. Module ``RELAX" handles the field redistribution triggered by both the primary and subsequently triggered instabilities. The whole avalanche, comprised of a seed, and subsequently triggered instabilities, is considered as one single flare. After complete relaxation, we further drive the system via the ``LOAD" module. There, a randomly-selected grid site receives a random magnetic field increment.


\subsection{``EXTRA": a nonlinear force-free extrapolation module}

The first step is the extrapolation of the photospheric magnetic fields.
As explained in \cite{dim09}, a physically meaningful treatment is the nonlinear force-free (NLFF)
field extrapolation. Our method of choice is based on the optimization technique introduced by
Wheatland et al. (2000) and further developed by T. Wiegelmann and collaborators (\cite{wie04, wie06, wie08}). This technique reconstructs force-free magnetic
fields from their boundary values by minimizing the Lorentz force and the divergence of the magnetic field vector in the extrapolation volume:

   \begin{equation}
      L=\int_{V}w(x,y,z)[|\vec{B}|^{-2}|(\vec{\nabla}\times\vec{B})\times\vec{B}|^{2}+|\nabla\cdot\vec{B}|^{2}]d^{3}x
   \end{equation}

In the above functional,  $w(x,y,z)$ is a weighting function and $V$ denotes the extrapolation volume.
A force-free state is reached when $L\rightarrow0$ for $w>0$. For $w(x,y,z)=1$
the magnetic field must be available on all 6 boundaries of our cubic box for the optimization algorithm to work.
However, photospheric vector magnetograms pertain only to the bottom boundary,
whereas the magnetic field vector on the top and lateral boundaries is unknown. The weighting function is thus used in order to
minimize the dependence of the interior solution from the unknown boundaries. In this study we introduce a buffer zone
of $10$ grid points expanding to the lateral and top boundaries of the computational box. We then choose $w(x,y,z)=1$ in
the inner domain and let $w$ drop to 0 with a cosine-profile in the buffer zone towards the lateral and top
boundaries of the computational box. This technique was first described by Wiegelmann (2004).

An additional useful attribute of Wiegelmann's NLFF field extrapolation code is the preprocessing option it offers. As the
photospheric magnetic field is in principle inconsistent with the force-free approximation, a preprocessing procedure was
developed by Wiegelmann et al. (2006) in order to drive photospheric fields closer to a NLFF field equilibrium. Preprocessing minimizes the forces and torques in the system, thus satisfying the force-free requirements more closely.

Although NLFF extrapolation methods have been greatly improved over recent years, such models still include numerous uncertainties (\cite{sch09}). Additional constraints stem from the measurements (signal-to-noise ratio, inadequate resolution of the $180^{o}$ ambiguity) or from physical origins (variation in the line formation height, the non-force-free nature of the photospheric vector magnetograms), which are not adequately handled in the course of the extrapolation. Such uncertainties are unavoidably conveyed to our simulations.

%


\subsection{``DISCO": a module to identify magnetic-field instabilities}

We assume that instabilities occur if the magnetic field stress exceeds a critical threshold.
For every site $\vec{r}$ within our grid we calculate the magnetic field stress $G_{av}(\vec{r})$ as\\

           $G_{av}(\vec{r})=|\vec{G_{av}}(\vec{r})|$\\

           where\\

           $\vec{G_{av}}(\vec{r})=\vec{B}(\vec{r})-\frac{1}{nn}\sum_{nn}\vec{B}_{nn}(\vec{r})$\\














           In the above definitions $nn$ is the number of nearest neighbors for each site $\vec{r}$ and $\vec{B}_{nn}(\vec{r})$ is the magnetic field vector of these neighbors.
           Depending on the location of each site within the volume, the number of nearest neighbors $nn$ can be
           $nn=3,4,5,6$. The physical reason
           for selecting this criterion lies in the fact that large magnetic stresses favor magnetic reconnection in three dimensions, even in the absence
           of null points (\cite{pri03}).





Mathematically, it can be shown that the selection of  $G_{av}$ as the critical quantity in our model
relates to the diffusive term of the induction equation (see \cite{isl98} for a detailed discussion).
Let us write the induction equation in the form:

   \begin{equation}
      \frac{\partial{\vec{B}}}{\partial{t}}= \vec{\nabla}\times(\vec{V}\times\vec{B})+\eta{{\nabla}}^{2}{\vec{B}}
\end{equation}

where $\vec{V}$ is the plasma velocity and $\eta$ is the resistivity.
The Laplacian of the magnetic field $\nabla^2\vec{B}(\vec{r})$ can be written as follows:\\

$\nabla^2\vec{B}(\vec{r})=(\nabla^2 B_x)\hat{\textbf{i}}+(\nabla^2 B_y)\hat{\textbf{j}}+(\nabla^2 B_z)\hat{\textbf{k}}$,\\

where $\vec{r}=(i,j,k)$. Letting $m\equiv{x,y,z}$ we obtain\\

$\frac{\partial^{2}B_{m}(\vec{r})}{\partial{x^2}}\simeq{\frac{1}{\Delta{x^2}}(B_{m_{i+1,j,k}}+B_{m_{i-1,j,k}}-2B_{m_{i,j,k}})}$\\

$\frac{\partial^{2}B_{m}(\vec{r})}{\partial{y^2}}\simeq{\frac{1}{\Delta{y^2}}(B_{m_{i,j+1,k}}+B_{m_{i,j-1,k}}-2B_{m_{i,j,k}})}$\\

$\frac{\partial^{2}B_{m}(\vec{r})}{\partial{z^2}}\simeq{\frac{1}{\Delta{z^2}}(B_{m_{i,j,k+1}}+B_{m_{i,j,k-1}}-2B_{m_{i,j,k}})}$\\

adopting a central finite-difference scheme and using the general case of a grid point having 6 nearest neighbors ($nn=6$). Further assuming
$\Delta{x}=\Delta{y}=\Delta{z}=1$ (the grid-size) we have:\\

$\nabla^2{B_{m}({\vec{r}})}=\frac{\partial^{2}B_{m}(\vec{r})}{\partial{x^2}}+\frac{\partial^{2}B_{m}(\vec{r})}{\partial{y^2}}+\frac{\partial^{2}B_{m}(\vec{r})}{\partial{z^2}}\simeq{\sum_{nn}B_{m_{nn}}-nn B_{m_{i,j,k}}}$\\

which yields $\nabla^{2}\vec{B}(\vec{r})$ as follows:\\

$\nabla^{2}\vec{B}(\vec{r})\simeq{\sum_{nn}\vec{B_{nn}}(\vec{r})-nn\vec{B}(\vec{r})}$\\

From the definition of the critical quantity $\vec{G_{av}}(\vec{r})$ it follows that:

\begin{equation}
\nabla^2\vec{B}(\vec{r})\simeq-nn\vec{G_{av}}(\vec{r})
\end{equation}

Therefore the critical quantity $G_{av}(\vec{r})$ relates directly to the Laplacian ${{\nabla}}^{2}{\vec{B}}$.
The resistivity in the solar corona is almost zero everywhere except in regions where the discontinuities (and the local currents) reach a critical value. In these regions current-driven instabilities will enhance the resistivity by many orders of magnitude and the second term in equation (2) will become dominant.
The convective term $\vec{\nabla}\times(\vec{V}\times\vec{B})$ of equation (2) will be further discussed in section 3.4,
where the driving module ``LOAD" is described.

There are several ways to determine the threshold value for the critical quantity,
above which a site is considered unstable:

\begin{enumerate}
\item We apply a histogram method, by constructing the histogram of the $G_{av}$
values in our grid. We then fit a Gaussian to this histogram and define the
threshold $G_{cr}$ as the field stress value, above which
the histogram deviates from the Gaussian.

\item We define the threshold value $(G_{cr})$ for the whole grid, as the maximum
$G_{av_{max}}$ value throughout our volume, slightly decreased:\\
$G_{cr}=G_{av_{max}}(1-s)$\\
where $s<<1$

\item We define the threshold value $(G_{cr}(z))$ per height $z$ , as the maximum
$G_{av_{max}}$ value for each specific height, slightly decreased:\\
$G_{cr}(z)=G_{av_{max}}(z)(1-s)$\\
where $s<<1$

\item We define the threshold value as a function of height $z$, e.g.:\\
$G_{cr}(z)=G_{av_{max}}(1-s)\exp{(-z)}$\\
where $s<<1$
\end{enumerate}

Here we present the results produced by the first (histogram) method, which yielded
$G_{cr}=10 G$ for our sample, and shortly refer to the other threshold alternatives in Sect.4.
Every site $\vec{r}=(i,j,k)$ for which the inequality $G_{av_{i,j,k}}\geq{G_{cr}}$ is satisfied, is considered
unstable and undergoes magnetic field restructuring under the rules implemented in the ``RELAX" module. Notice that, given the definition of the critical threshold, instabilities sometimes occur even from the first iteration, after constructing the NLFF fields. This, however, does not incur any qualitative impact on the evolution of the system toward SOC, or the statistical results of the simulation.



\subsection{``RELAX": a redistribution module for magnetic energy}

In case the instability criterion $G_{av_{i,j,k}}\geq{G_{cr}}$ is met for a specific site $i,j,k$, then the
vicinity of the unstable location undergoes a field restructuring, which follows the rules of
\cite{luh91}:

\begin{equation}
\vec{B}^+(\vec{r})\rightarrow{\vec{B}(\vec{r})-\frac{6}{7}\vec{G}_{av}(\vec{r})}
\end{equation}

\begin{equation}
\vec{B}^+_{nn}(\vec{r})\rightarrow{\vec{B}_{nn}(\vec{r})+\frac{1}{7}\vec{G}_{av}(\vec{r})}
\end{equation}

where the superscript $+$ denotes the field components after the redistribution.


At this point it is important to investigate whether the redistribution rules as defined here violate basic physical laws, such as the zero-divergence requirement for the magnetic field. The initial magnetic configuration satisfies approximately the condition $\vec{\nabla}\cdot\vec{B}=0$, as the field has been reconstructed using an NLFF field extrapolation. The question is whether the magnetic field after the redistribution imposed by rules (4)-(5) still satisfies the same demand ($\vec{\nabla}\cdot\vec{B^{+}}=0$). Taking the divergence of $\vec{B^{+}}(\vec{r})$ and its neighbors $\vec{B^{+}_{nn}}(\vec{r})$ we find from
relations (4) and (5) respectively:\\

$\vec{\nabla}\cdot{{\vec{B}}^{+}(\vec{r})}\simeq{\vec{\nabla}\cdot{{\vec{B}}(\vec{r})}-\frac{6}{7}\vec{\nabla}\cdot{{\vec{G}}_{av}(\vec{r})}}$\\

$\vec{\nabla}\cdot{{\vec{B}}^{+}_{nn}(\vec{r})}\simeq{\vec{\nabla}\cdot{{\vec{B}_{nn}}(\vec{r})}+\frac{1}{7}\vec{\nabla}\cdot{{\vec{G}}_{av}(\vec{r})}}$\\

From the definition of ${\vec{G}}_{av}(\vec{r})$ we now have:\\

$\vec{\nabla}\cdot{{\vec{G}}_{av}(\vec{r})}=\vec{\nabla}\cdot{{\vec{B}}(\vec{r})}-\frac{1}{nn}{\vec{\nabla}\cdot{{\vec{B}_{nn}}(\vec{r})}}$\\

Substituting this to the above we find:\\

\begin{equation}
\vec{\nabla}\cdot{{\vec{B}}^{+}(\vec{r})}\simeq{\frac{1}{7}\vec{\nabla}\cdot{{\vec{B}}(\vec{r})}-\frac{1}{7nn}{\vec{\nabla}\cdot{{\vec{B}_{nn}}(\vec{r})}}}
\end{equation}
\begin{equation}
\vec{\nabla}\cdot{{\vec{B}}^{+}_{nn}(\vec{r})}\simeq{\frac{1}{7}\vec{\nabla}\cdot{{\vec{B}}(\vec{r})}+\frac{1}{7nn}{\vec{\nabla}\cdot{{\vec{B}_{nn}}(\vec{r})}}}
\end{equation}

As $\vec{\nabla}\cdot{{\vec{B}}(\vec{r})}\simeq{\vec{\nabla}\cdot{{\vec{B}_{nn}}(\vec{r})}}\simeq{0}$ from our first iteration (extrapolated fields), we find $\vec{\nabla}\cdot{{\vec{B}}^{+}(\vec{r})}\simeq{\vec{\nabla}\cdot{{\vec{B}}^{+}_{nn}(\vec{r})}}\simeq{0}$. Thus, the redistribution of the magnetic field maintains the divergence-free condition.

Isliker et al. (1998) showed that the redistribution rules (4) and (5) implement local diffusion and after redistribution, $\vec{G}^+_{av}(\vec{r})\simeq{0}$,
so the instability at location $\vec{r}$ has been relaxed.


\subsection{``LOAD": the driver}

After the system is completely relaxed, we introduce a driving mechanism that adds a magnetic field increment
$\vec{\delta{B}(\vec{r})}$
at one randomly selected site
$\vec{r}$ within our grid. The driving process complies with the following conditions:

\begin{enumerate}

	 \item
  \begin{equation}
      \vec{B(\vec{r})\cdot\vec{\delta{B(\vec{r})}}}=0
   \end{equation}
    This condition implies that the magnetic field increment is always perpendicular to the existing magnetic field $\vec{B(\vec{r}})$ at the randomly selected site $\vec{r}$. Figure 1 provides a sketch of the suggested situation, depicting the directions of the plasma velocity $\vec{V}$, the magnetic field $\vec{B}$ and the perpendicular magnetic field increment $\vec{\delta{B}}$. We note that the condition described by Eq. (8) is compatible with two physical scenarios: (a) that Alfven waves may have been excited locally, or (b) that, according to the convective term $\vec{\nabla}\times(\vec{V}\times\vec{B})$ of the induction Eq. (2), a magnetized plasma upflow occurs in the AR, out from the photosphere.

   \begin{figure}
   \centering
  \includegraphics[width=0.5\textwidth]{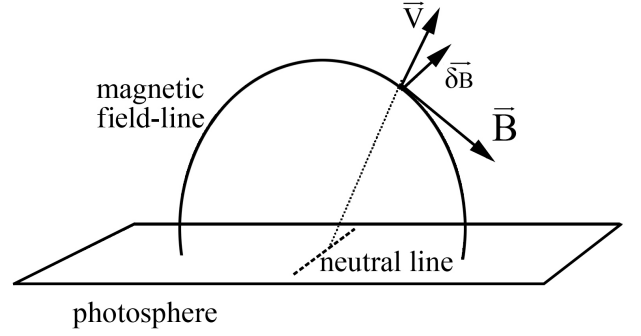}
   \caption{Typical configuration of a magnetic loop anchored in the photosphere.
   The magnetic field vector $\vec{B}$
   is perpendicular to an assumed plasma outflow velocity $\vec{V}$.
   The model driver requires that
   the magnetic field increments $\vec{\delta{B}}$ are always perpendicular to the existing magnetic field $\vec{B}$.}
              \label{FigGam}%
    \end{figure}


  \item
   \begin{equation}
      \frac{|\vec{\delta{B(\vec{r})}}|}{\vec{|B(\vec{r})|}}=\epsilon, \epsilon<1
   \end{equation}
   This is a typical condition known to allow the system reach the SOC state, without the latter being influenced by the loading process (\cite{bak87}). As also shown by \cite{luh91}, decreasing the driving rate by making the magnetic field increments even smaller, increases the average time between subsequent events.
   For the results presented here we have used a fixed $\epsilon=0.3$.

  \item
   \begin{equation}
    \vec{\nabla\cdot{(\vec{{B(\vec{r})}}+\vec{\delta{B(\vec{r})}})}}=0
   \end{equation}
 This condition should guarantee that the divergence of the magnetic field is approximately kept to zero during the loading process, as it was done during the redistribution of the magnetic field (RELAX module).
 In order to implement the condition, a first-order, left finite-difference scheme is used. In this way , however, condition (10)
 does not provide adequate guarantee for a divergence-free magnetic field in the selected site's vicinity.
 This is a known problem, which can be tackled by working with the vector potential $\vec{A}$, with $\nabla\times{{\vec{A}}}=\vec{B}$, instead of the magnetic field $\vec{B}$ directly (see e.g. \cite{luh93}, \cite{gal96}, Isliker et al. (2000), Isliker et al. (2001)). Because our study uses observed vector magnetograms as initial conditions, we naturally work with the known magnetic fields, rather than the unknown vector potential. Thus, equation (10) only provides a low-order approximation towards a divergence-free magnetic field. To monitor how effective condition (10) is, we introduce a ``Weighted Nabla Dot B" ($WNDB$) monitoring parameter, as follows:\\

 $WNDB= \frac{|\vec{\nabla\cdot{\vec{{B}}}}|}{\sqrt{3}\sqrt{(\frac{\partial{B_{x}}}{\partial{x}})^{2} + (\frac{\partial{B_{y}}}{\partial{y}})^{2} + (\frac{\partial{B_{z}}}{\partial{z}})^{2}}}$\\




 By definition, $WNDB$ is a dimensionless quantity, lying in the range $0\leq{WNDB}\leq{1}$.
 Monitoring $WNDB$ during our simulation will provide evidence on whether condition (10) can be
 considered adequate for keeping the magnetic field within our volume approximately divergence-free.
 In the following, we will tolerate a departure from zero of up to $20\%$ (${WNDB}\leq{0.2}$) for a still
 a roughly divergence-free magnetic field.

\end{enumerate}


\subsection{Model parameters}

Linking the above-mentioned modules in one consistent simulation, we construct a relatively simple algorithm and we monitor the flare duration, the peak energy and the total energy, the distribution functions of which we intend to compare against those of observational data. If an instability is identified (DISCO) - either in the original magnetic configuration generated by the initial extrapolated magnetogram (EXTRA) or due to an increment $\vec{\delta{B}}$ randomly added at a grid point (LOAD) - the possible chain of instabilities that follows is left to completely relax (RELAX) before an additional magnetic field increment is randomly placed (LOAD), possibly causing a new instability. This rule takes into account the observational fact that the lifetime of a flare is much
smaller than the evolution timescale of an AR. Successive grid scans may be required for an instability to be completely relaxed. Each scanning corresponds to one \textbf{timestep}. Therefore, the relaxation of an event may be accomplished in more than one timesteps. Each loading event according to the equation set (8)-(10) commences a new \textbf{iteration}. The \textbf{duration of an event} is defined as the total number of timesteps the event lasted, from its onset until its complete relaxation. The accumulated released energy during the event provides the \textbf{total energy of an event}, whereas the peak during an event yields the \textbf{peak energy/luminosity of an event}.

The simulation results presented in the next section have been performed using a $32 x 32 x 32$ cubic grid with ``open" boundaries in the relaxation events (see Isliker et al. (2001) for a detailed discussion on open boundary conditions). Each simulation is driven for $3 x 10^{5}$ iterations, which equals the times that LOAD module is being called during the simulation. This mechanism allows the production of multiple subsequent flares in each AR. In all cases the critical threshold $G_{cr}$ was kept fixed and equal to $10 G$.


\section{Results}

 Applying our flare simulation model to our 11-event-database, we find that in all cases the simulated ARs reached the SOC state. An indication on whether and when the SOC state has been reached is obtained by monitoring the quantity $\bar{G_{av}}$, namely, the volume average of the critical quantity $G_{av}$. During the continuous driving of the system and the subsequently generated avalanches, $\bar{G_{av}}$ increases gradually. When the system reaches the SOC state, $\bar{G_{av}}$ stabilizes around a value which depends on the system's characteristics. For the loading method used in our model (new magnetic field increments are only added when a previously triggered avalanche has decayed), the value around which $\bar{G_{av}}$ stabilizes is slightly lower than the threshold value $G_{cr}$. A second indication that the system has reached the SOC state is that the total energy of the system tends to an asymptotic value. This is because SOC is a statistically stationary state. Figure 2 shows the $\bar{G_{av}}$ value over $3x10^{5}$ timesteps for $AR10570$. $\bar{G_{av}}$ is constantly increasing up to timestep $1.4x10^{5}$, thereafter stabilizing at $\sim{9.80G}\lesssim{G_{cr}}=10G$. Similarly, Fig. 3 shows the logarithm of the total magnetic energy throughout the volume $E_{totaft}$ after each scan of the grid for possible redistributions. Following $\bar{G_{av}}$, $E_{totaft}$ increases until a stable state is reached. The stabilization of both $\bar{G_{av}}$ and $E_{totaft}$ is a solid indication that SOC has been reached for $AR10570$. The same behavior is seen for all ARs included in our sample.

        \begin{figure}
   \centering
  \includegraphics[width=0.5\textwidth]{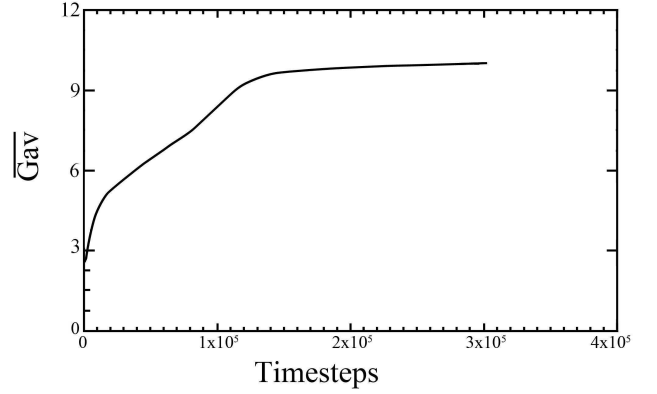}
   \caption{Average Laplacian $\bar{G_{av}}$ over the grid for $3x10^{5}$ timesteps for $AR10570$.
   $\bar{G_{av}}$ increases gradually until timestep $1.4x10^{5}$, after which the SOC state is reached.}
             \label{FigGam}%
    \end{figure}

       \begin{figure}
   \centering
  \includegraphics[width=0.5\textwidth]{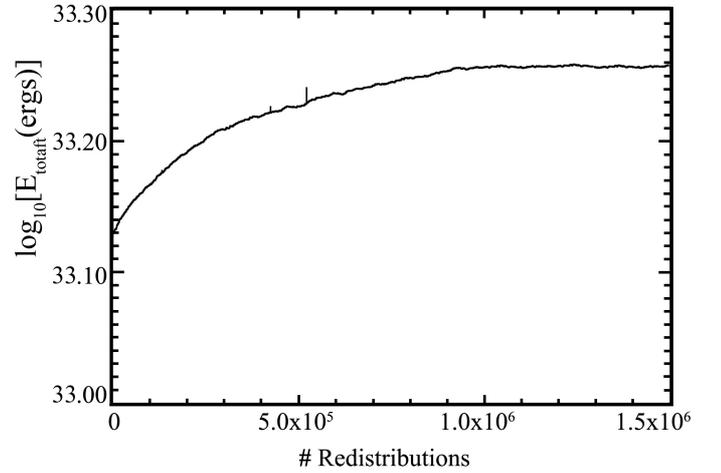}
   \caption{Diagram of $\log_{10}(E_{totaft})$ after each redistribution for $AR10570$.
   Like $\bar{G_{av}}$, $E_{totaft}$ increases gradually until a stable state is reached.}
              \label{FigGam}%
    \end{figure}

SOC is generally characterized by intermittent transport events (avalanches), whose sizes range from very small (a single neighborhood) up to comparable to the system size. Power-law frequency distributions describe the
parameters of these avalanches. It is thus reasonable to expect that since all 11 ARs
in our sample have reached the SOC state under the imposed driving rules, they should all
produce distribution functions for the flare duration, peak energy and total energy, which either follow pure power laws or
functions including a power law part (e.g. power laws with exponential rollover).
The functions tested against the model results for all flare parameters (flare duration, peak energy, total energy)
were single power laws, double power laws, power laws with exponential rollover, and exponential functions.
In order to define the best-fitting function per case, we made least square fits and performed chi-square goodness-of-fit tests.

%
\begin{table*}
\caption{Power-law indices and respective chi-square probabilities derived by the best-fit functions for the flare duration
for the 11 ARs comprising our sample.
}             
\label{table:1}      
\centering                          
\begin{tabular}{c c c c c c c}        
\hline                
\multicolumn{7}{c}{FLARE DURATION (model)}\\
\hline                
\multicolumn{1}{c} {}& \multicolumn{4}{c}{Double Power Law fit} & \multicolumn{2}{c}{Power Law with }\\
\hline                
\multicolumn{1}{c} {} & \multicolumn{2}{c}{Flat PL} & \multicolumn{2}{c}{Steep PL}& \multicolumn{2}{c}{Exponential Rollover}\\
\hline                
AR  & PL Index   & $Probability$ & PL Index   & $Probability$ &   PL Index   & $Probability$    \\    
\hline                        
9415 &  ... &	...	&  ... &	...	    &$-1.42\pm{0.18}$	 & 	$0.95$	      \\
9635 &  $-2.29\pm{0.19}$& $0.96$&$-5.28\pm{0.42}$ &	$0.95$	 &    ...	 & ...       \\
9661 &  ...	&... &  ...&	...	 &   $-0.26\pm{0.05}$	 & 	$0.94$     \\
9684 &  ...	&... &  ... &	...	 &    $-0.91\pm{0.09}$	 & 	$0.95$    \\
9845 &  ...	&...	&  ... &	... &    $-1.12\pm{0.06}$	 & 	$0.95$	    \\
10050&  $-1.80\pm{0.18}$	& $0.98$	&  $-4.03\pm{0.29}$ &	$0.95$	 &    ...	 & ...	       \\
10247&  ...	&...	&  ... &	...	 &    $-1.27\pm{0.07}$	 & 	$0.95$   \\
10306&  ...	&... &  ... &	...	 &    $-1.27\pm{0.09}$	 & 	$0.94$  \\
10323&  ...	&...	&  ... &	...	 &    $-0.98\pm{0.05}$	 & 	$0.95$   \\
10488&  $-1.60\pm{0.16}$	&$0.94$	&  $-3.64\pm{0.19}$ &	$0.94$	 &    ... & 	...   \\
10570&  ...	&...&  ... &	...	 &    $-0.83\pm{0.07}$	 & 	$0.95$     \\
\hline
MEAN & $-1.90$ & & $-4.32$&  & $-1.01$ & \\
\hline
$\sigma^2$ & $0.35$ & & $0.86$&  & $0.36$ & \\
\hline
\end{tabular}
\end{table*}

\begin{figure*}
  \centering
  \includegraphics[width=0.5\textwidth]{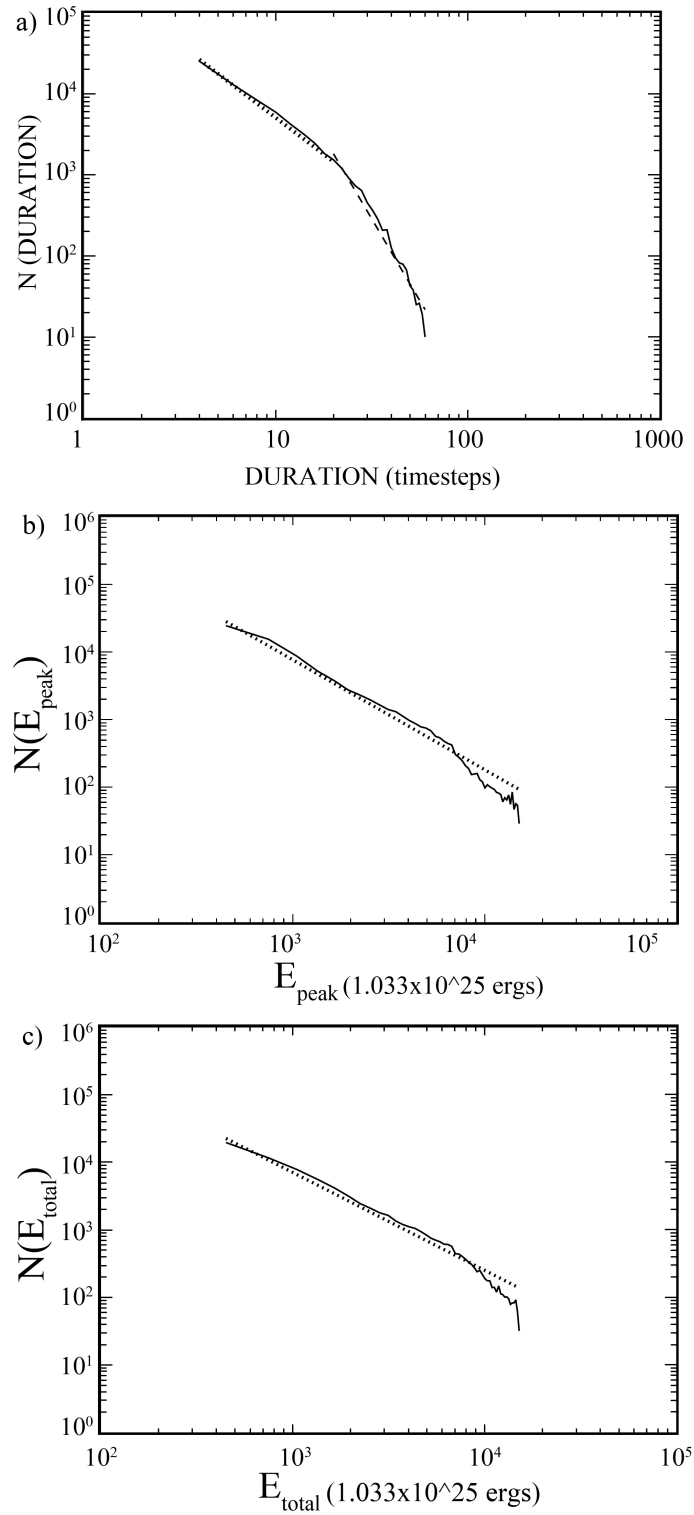}
   \caption{Distribution functions for the event duration (Fig. 4a), the peak energy $E_{peak}$ (Fig. 4b) and
   the total energy $E_{total}$ (Fig. 4c) for $AR10050$. The energies in b) and c) are calculated in physical units ($ergs$).}
              \label{FigGam}%
    \end{figure*}
    
    \begin{figure*}
  \centering
  \includegraphics[width=0.5\textwidth]{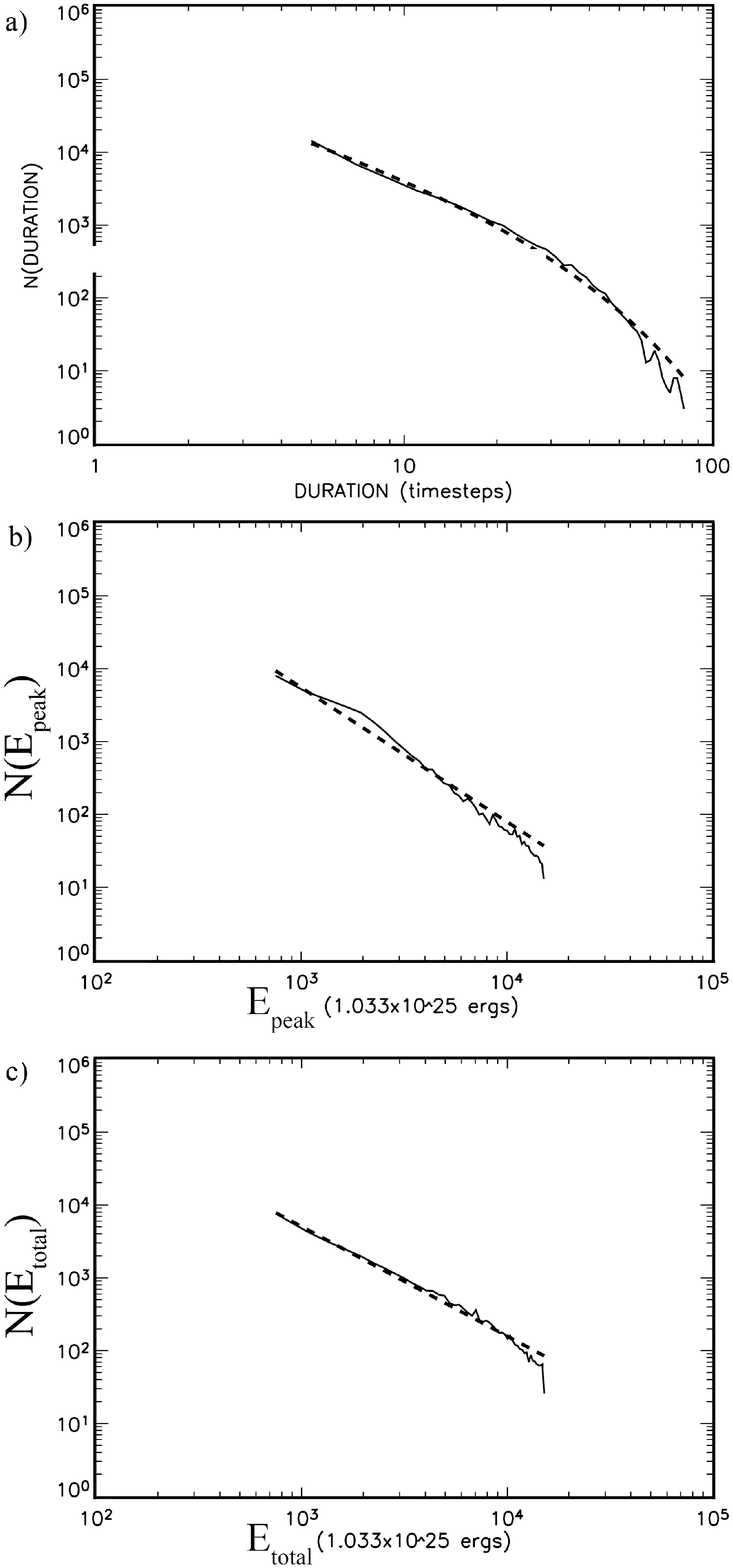}
  \caption{Same as Fig. 4 for $AR9415$.}
   \label{FigGam}%
    \end{figure*}

Figures 4 and 5 are typical examples of our general results. Figure 4 depicts the distribution functions of duration (Fig. 4a), peak energy (Fig. 4b) and total energy (Fig. 4c)) for $AR10050$. The duration distribution follows a double power law with index $-1.80\pm{0.18}$ for the flatter part and $-4.03\pm{0.29}$ for the steeper part. Both the peak and total energy distribution functions follow single power laws with indices $-1.63\pm{0.15}$ and $-1.45\pm{0.13}$, respectively. Figure 5 depicts the distribution functions of duration (Fig. 5a), peak energy (Fig. 5b) and total energy (Fig. 5c) for $AR9415$. Here the duration distribution follows a power law with an exponential rollover. The power law index is $-1.42\pm{0.18}$. The peak and total energy distribution functions follow again single power laws with indices $-1.84\pm{0.18}$ and $-1.50\pm{0.13}$, respectively.



From the above, flare-duration distributions appear to be best-fitted either by power laws or by power laws with exponential rollovers.
By comparing the chi-square values and respective probabilities for single power laws, double power laws and power laws with exponential rollover respectively, it is concluded that 3 of our ARs follow double power laws ($AR9635$, $AR10050$, $AR10488$), while the rest are best fitted by power laws with exponential rollovers. These indices are summarized in Table 1, along with the respective probabilities. The values shown in Table 1 refer to the fitting achieved against the entire distribution function in all cases (all bins included). Single power laws fail to describe the model duration distribution functions in all cases, whereas exponential functions only fit the tail of the generated model curves. In cases where a double power law is the best fit ($AR9635$, $AR10050$, $AR10488$), the mean index for the flat power law is $-1.90$, whereas the mean index for the steep power law is $-4.32$. When power laws with exponential rollover are best fitting (remaining ARs), then the mean value for the power law index is $-1.01$. Standard deviations ($\sigma^2$) to these mean values are given in the last row of this table.

Although single power laws are not the optimum functions to fit the modeled flare duration, they are undoubtedly the best-fitting theoretical functions for the peak energy and the total flare energy. As shown in Table 2 for the peak flare energy and in Table 3 for the total flare energy, the average value for $E_{peak}$ is $-1.80$, whereas the average index value for $E_{total}$ is $-1.57$. The standard deviation ($\sigma^2$) of these mean values is given in the last row of these tables.

\begin{table}
\caption{Power-law indices and respective chi-square probabilities derived by fitting a single power law to the peak flare energy
for the 11 ARs comprising our sample.
}             
\label{table:1}      
\centering                          
\begin{tabular}{c c c }        
\hline                
\multicolumn{3}{c}{FLARE PEAK ENERGY (model)}\\
\hline                
\multicolumn{3}{c}{Single Power Law fit}\\
\hline                
AR  & PL Index   & $Probability$    \\    
\hline                        
9415 &  $-1.84\pm{0.18}$	&$0.95$	   \\
9635 &  $-2.62\pm{0.17}$	&$0.97$	   \\
9661 &  $-1.42\pm{0.15}$	&$0.98$	   \\
9684 &  $-1.70\pm{0.17}$	&$0.97$	   \\
9845 &  $-1.85\pm{0.12}$	&$0.95$	    \\
10050&  $-1.63\pm{0.15}$	&$0.95$	    \\
10247&  $-2.15\pm{0.12}$	&$0.98$	   \\
10306&  $-1.61\pm{0.16}$	&$0.97$	   \\
10323&  $-1.72\pm{0.17}$	&$0.97$	  \\
10488&  $-1.59\pm{0.14}$	&$0.95$	    \\
10570&  $-1.63\pm{0.15}$	&$0.98$	   \\
\hline                                   
MEAN &  $-1.80$ 	&   \multicolumn{1}{c}{} \\
\hline                                   
$\sigma^2$ &  $0.33$ 	&   \multicolumn{1}{c}{} \\
\hline                                   
\end{tabular}
\end{table}
%

%
\begin{table}
\caption{Power-law indices and respective chi-square probabilities derived by fitting a single power law to the total flare energy
for the 11 ARs comprising our sample.
}             
\label{table:1}      
\centering                          
\begin{tabular}{c c c }        
\hline                
\multicolumn{3}{c}{FLARE TOTAL ENERGY (model)}\\
\hline                
\multicolumn{3}{c}{Single Power Law fit}\\
\hline                
AR  & PL Index   & $Probability$    \\    
\hline                        
9415 &  $-1.50\pm{0.13}$	& $0.95$	     \\
9635 &  $-2.22\pm{0.19}$	& $0.98$  \\
9661 &  $-1.27\pm{0.05}$	& $0.99$	   \\
9684 &  $-1.43\pm{0.07}$	& $0.99$	  \\
9845 &  $-1.69\pm{0.17}$	& $0.95$	  \\
10050&  $-1.45\pm{0.13}$	& $0.95$    \\
10247&  $-1.89\pm{0.17}$	& $0.98$  \\
10306&  $-1.23\pm{0.08}$	& $0.99$	  \\
10323&  $-1.45\pm{0.16}$	& $0.98$	  \\
10488&  $-1.54\pm{0.13}$	& $0.95$	    \\
10570&  $-1.45\pm{0.08}$	& $0.99$	  \\
\hline                                   
MEAN &  -1.56 	&   \multicolumn{1}{c}{} \\
\hline                                   
$\sigma^2$ &  0.28 	&   \multicolumn{1}{c}{} \\
\hline                                   
\end{tabular}
\end{table}

Figure 6 illustrates
the magnetic energy released $E_{rel}$ for a specific period of $10000$ timesteps after SOC has been reached for $AR10570$.
As the added driver increments $\vec{\delta{B}}$ assume small and random values, the waiting time from one flaring event to another varies. Figures 7 and 8 show a 3d representation of the emerging magnetic discontinuities during a large and a smaller avalanche, respectively, simulated for $AR10247$ after SOC has been reached. In the former case, the avalanche during its early stages generates 140 discontinuities (Fig. 7a), evolves further (Fig. 7b) with 281 discontinuities, peaks (Fig. 7c) with 425 discontinuities, and decays (Fig. 7d, 7e, 7f) with 184, 51, and 18
discontinuities, respectively. The total event duration is 341 steps. The total duration of the smaller event (Fig. 8) is 90 steps. The event during its early stages generates 6
discontinuities (Fig. 8a), peaks (Fig. 8b) with 15 discontinuities, and decays (Fig. 8c, 8d) with 10 and 6 discontinuities, respectively.

   \begin{figure*}
   \centering
  \includegraphics[width=0.5\textwidth]{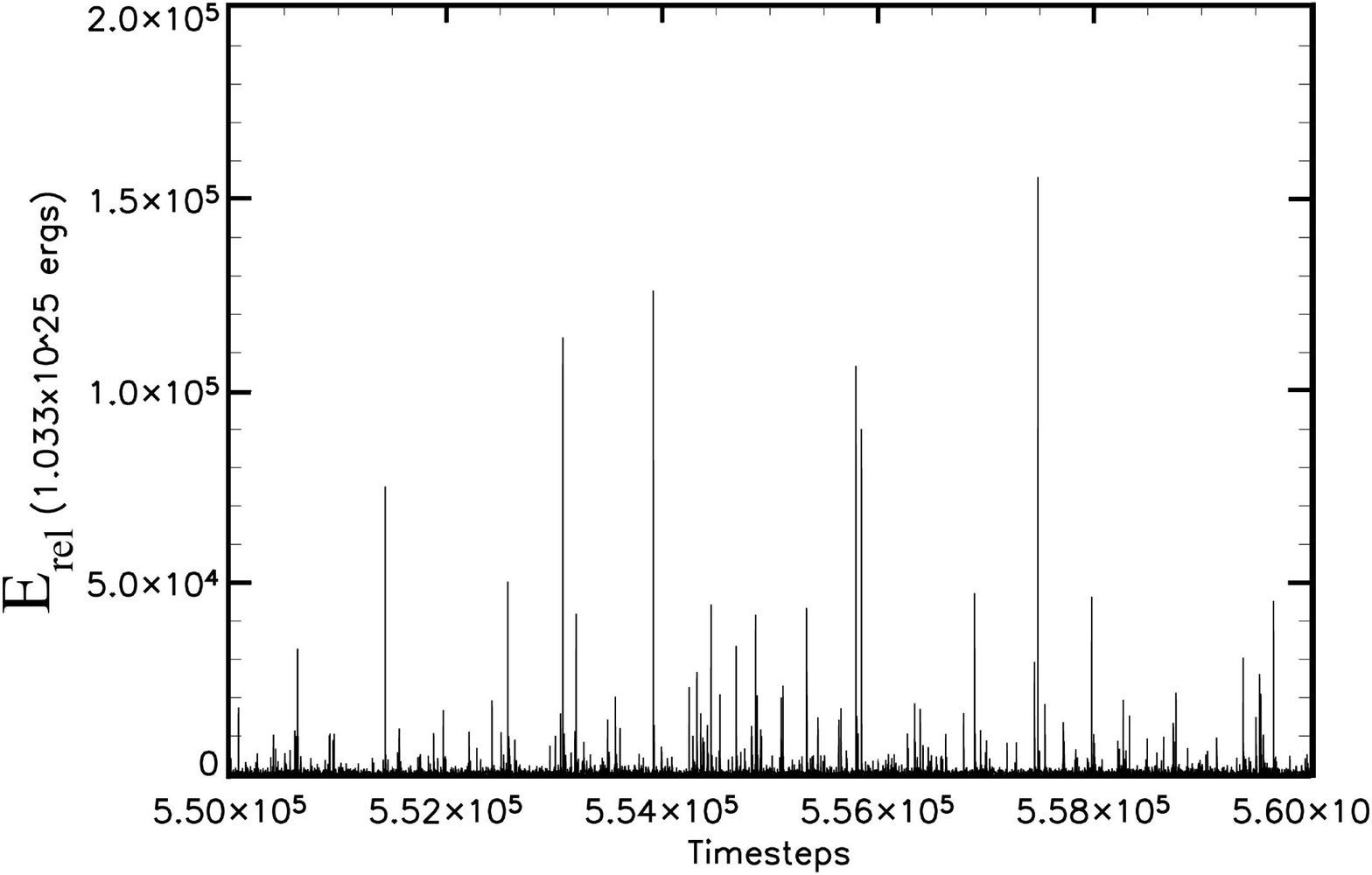}
   \caption{Time series of the total magnetic energy released $E_{rel}$ (in $1.033x10^{25} ergs$) from the simulated $AR10570$
   after the SOC state has been reached. The time series shown consists of $10000$ timesteps for better detail.}
              \label{FigGam}%
    \end{figure*}

   \begin{figure*}
   \centering
  \includegraphics[width=0.9\textwidth]{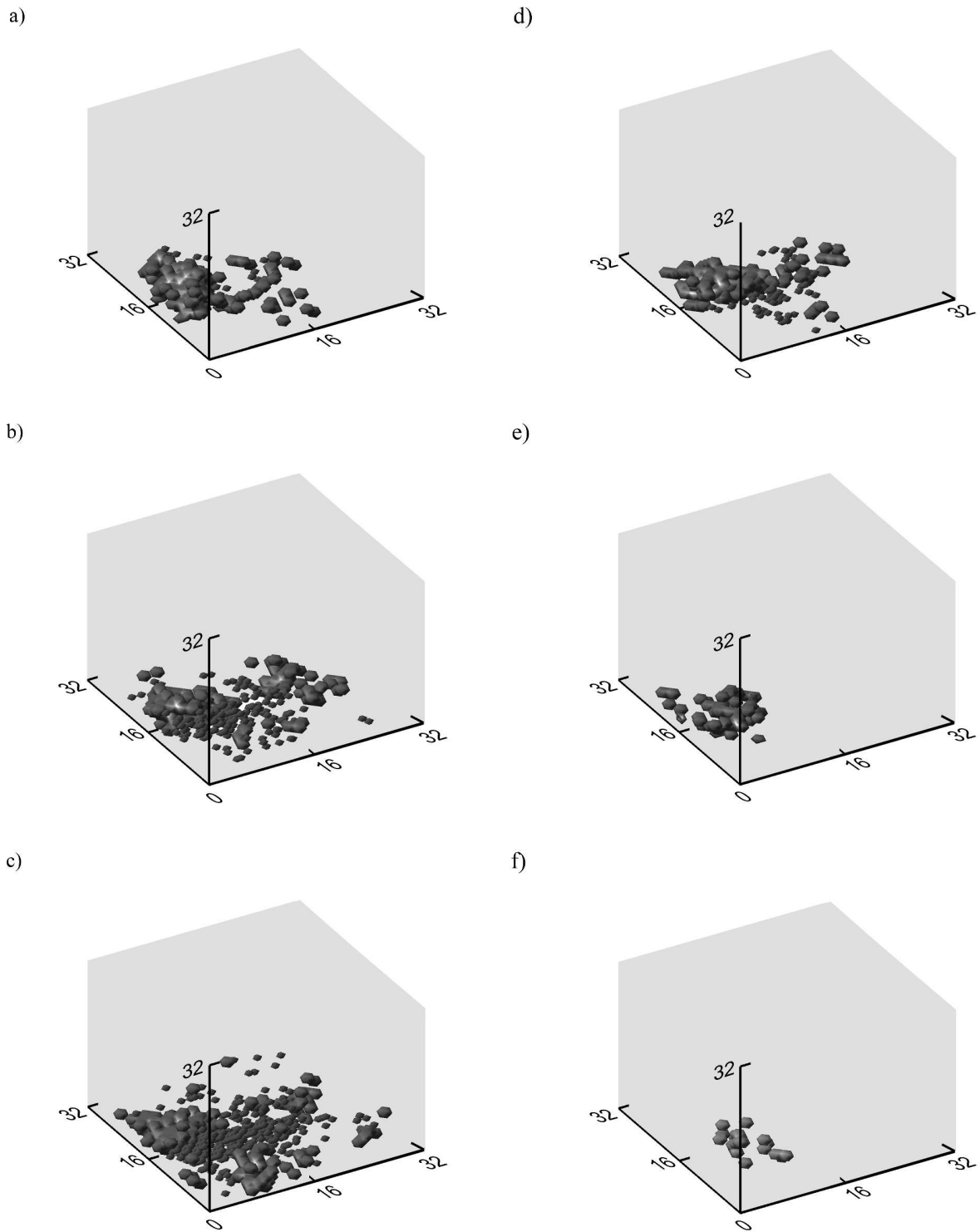}
   \caption{3d representation of the emerging magnetic discontinuities during an avalanche in $AR10247$.
   The total duration of this event is 341 steps. During the early stages, the avalanche generates numerous discontinuities (140 in a), evolves with 281 discontinuities (b), peaks with 425 discontinuities (c),
   and decays with 184 discontinuities (d), 51 discontinuities (e), and 18 discontinuities (f).}
              \label{FigGam}%
    \end{figure*}

     \begin{figure*}
   \centering
  \includegraphics[width=0.9\textwidth]{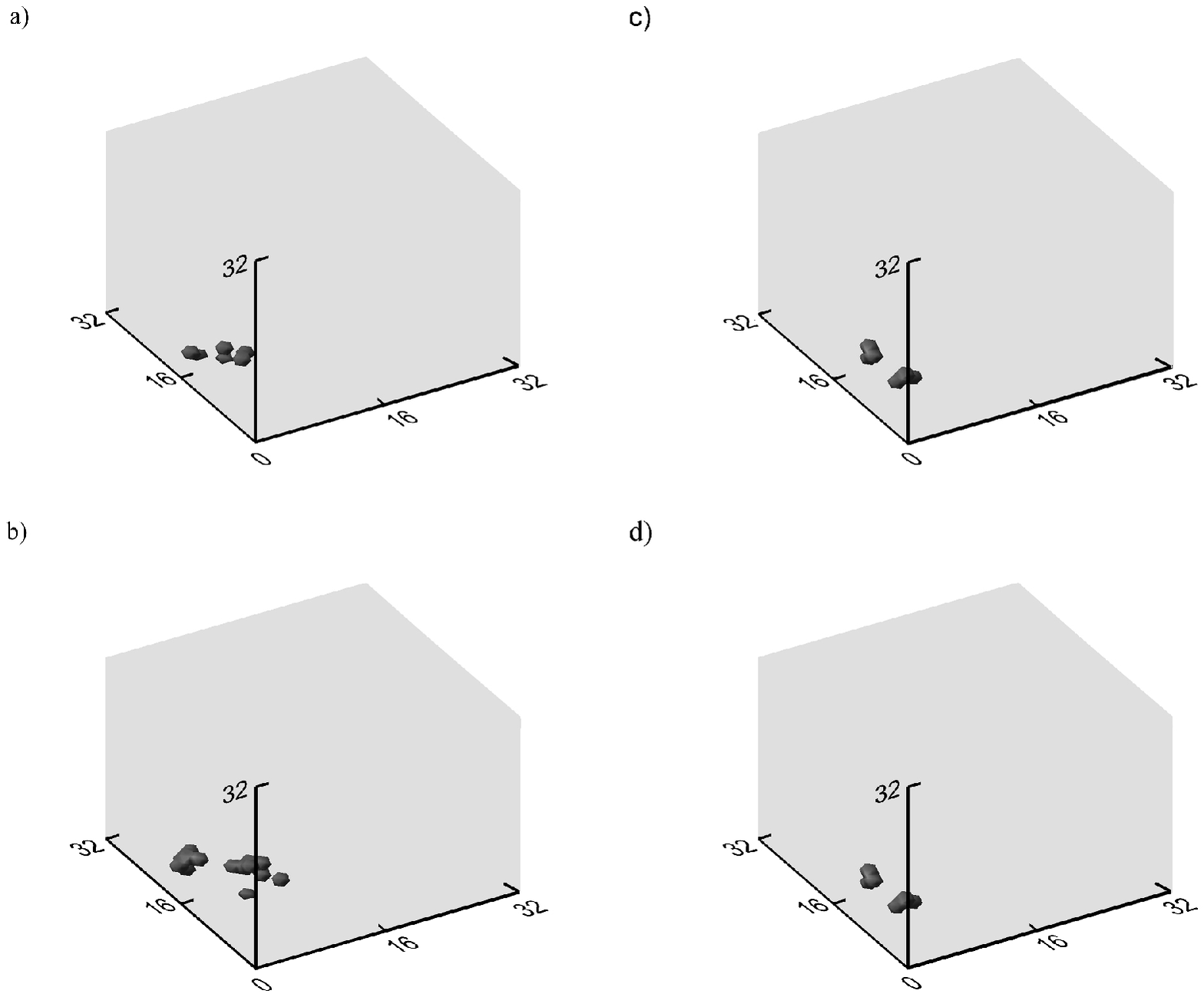}
   \caption{3d representation of the emerging magnetic discontinuities during an avalanche in $AR10247$.
    The total duration of this event is 90 steps. During the early stages, the event generates a small number or discontinuities (6 in a),
    peaks with 15 discontinuities (b), and decays with 10 discontinuities (c) and 6 discontinuities (d).}
              \label{FigGam}%
    \end{figure*}


Finally, it is interesting to investigate whether our model consistently reproduces the distribution functions of the flaring events actually observed in the ARs in our sample.
For our comparison we used the solar X-ray flare catalog from the GOES satellite (\url{http://www.ngdc.noaa.gov/stp/SOLAR/ftpsolarflares.html}, item 3). The flaring events recorded in this database lie in the class range $B-X$ and are summarized in Table 4 for each AR. However, to construct the distribution functions of flare
parameters we need sufficient statistics, reflected in large flare numbers. A single AR, regardless of flare productivity, is unlikely to provide these numbers. For this reason and
for the sake of comparison we have merged all observed flares in all studied ARs into a single flare sequence with a total of 154 events (sum of all flares in Table 4). Table 5 shows the statistical results of our analysis for flare durations. As flare duration we define the observed onset-to-end elapsed time. The best fitting function is not easily discernible in this case, as all candidate functions (single power law, double power law and power law with exponential rollover) fit the observational data fairly well. Figure 9 depicts the fit between the observed flare durations against a double power law (Fig. 9a) and a power law with exponential rollover (Fig. 9b). In the former case the calculated index is $-1.67\pm{0.09}$ for the smooth part and $-3.37\pm{0.25}$ for the steep part, whereas in the latter case the power law index yields $-1.28\pm{0.11}$. It is apparent that the dynamical range of the power-law in Fig. 9a is very limited, but it is shown here for comparison purposes. In order to achieve this comparison, we merge the model results of the separate runs per AR into one common database. Figure 10 depicts the fit between the merged model flare durations for all ARs in our sample
against a double power law (figure 10a) and a power law with exponential rollover (Fig. 10b). In the former case the calculated index is $-1.78\pm{0.27}$ for the smooth part and $-3.91\pm{0.42}$ for the steep part, whereas in the latter case the power law index yields $-1.13\pm{0.12}$. By comparing the results depicted in Fig. 9 (observational data) and Fig. 10 (merged model data), we conclude that the power law indices for the observational data are close
to our model's values for the double power law and power law with exponential rollover fits. This is not the case for the single power law fitting, which yields an index of $-1.70$ for the GOES data. The best agreement between the observed and the simulated flares is, therefore, achieved when the attempted fit is not a single power law, but either a double
power law or a power law with an exponential rollover. Table 6 is similar to Table 5, summarizing the indices resulting from the merged model data.

   \begin{figure*}
   \centering
   \includegraphics{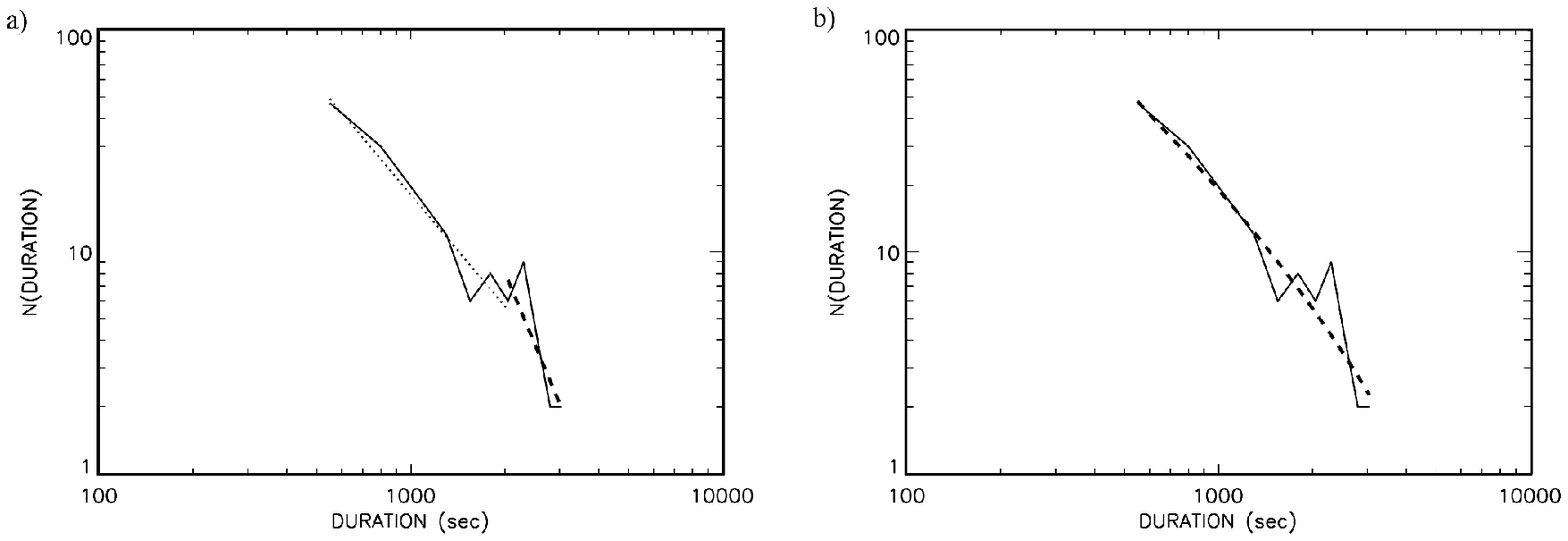}
   \caption{Observed distributions of GOES flare durations for all ARs in our sample. Fit is attempted using a double power law (a) and a power law with an exponential rollover (b).
The double power law fit yields an index equal to $-1.67\pm{0.09}$ for the flat part and $-3.37\pm{0.25}$ for the
steep part, whereas the fit with the power law and the exponential rollover yields a scaling index $-1.28\pm{0.11}$.}
              \label{FigGam}%
    \end{figure*}

   \begin{figure*}
   \centering
   \includegraphics{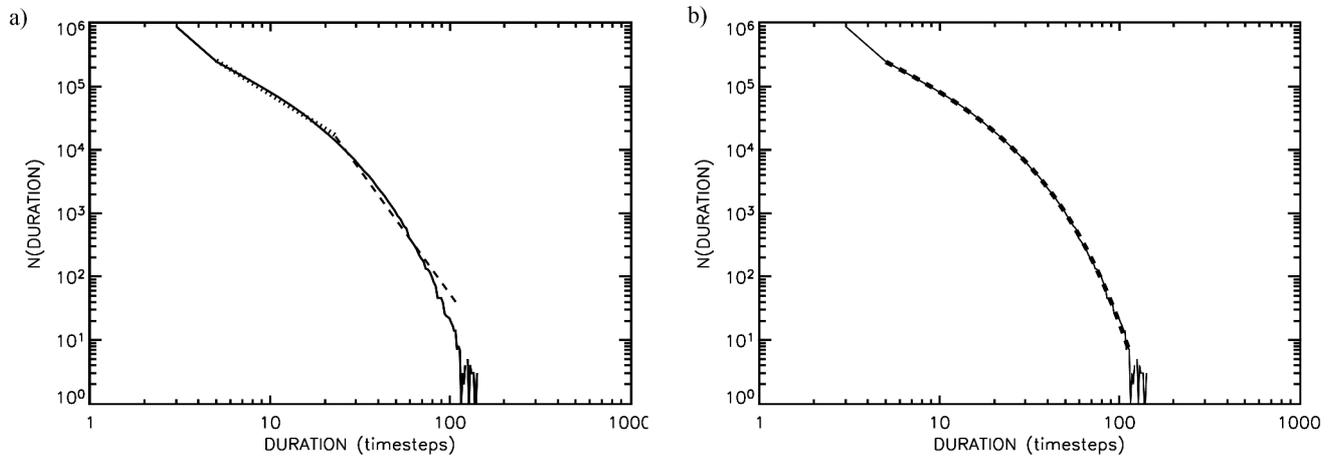}
   \caption{Distributions of simulated flare durations for all ARs in our sample. Fit is attempted using a double power law (a) and a power law with an exponential rollover (b).
The double power law fit yields an index equal to $-1.78\pm{0.27}$ for the flat part and $-3.91\pm{0.42}$ for the
steep part, whereas the fit with the power law and the exponential rollover yields a scaling index $-1.13\pm{0.12}$.}
              \label{FigGam}%
    \end{figure*}

\begin{table}
\caption{GOES X-ray data for the number and class of observed flares in the ARs used in our simulations.}             
\label{table:1}      
\centering                          
\begin{tabular}{c c c c c c}        
\hline                
AR  & B-Class  & C-Class & M-Class & X-Class & Total \\    
\hline                        
$9415$  &    03	&   16	 &	06    &	05   &	30 \\
$9635$      &    00	&   02   &	00    & 00   &	02 \\
$9661$  &    00	&   16   &	01    & 02   &	19 \\
$9684$      &    00	&   08   &	01    & 01   &	10 \\
$9845$      &    01 &   04   &	00    &	00   &	05 \\
$10050$ &    00	&   16   &	00    &	00   &	16 \\
$10247$     &    00	&   01   &	00    &	00   &	01 \\
$10306$     &    06	&   02   &	00    &	00   &	08 \\
$10323$     &    00	&   05   &	00    &	00   &	05 \\
$10488$ &    00	&   17   &	07    &	02   &	26 \\
$10570$ &    17	&   14   &	01    &	00   &	32 \\
\hline                                   
\end{tabular}
\end{table}
%

%
\begin{table*}
\caption{Power law indices and respective chi-square probabilities derived by fitting several functions
to the flare
durations
derived from the merged GOES observational data for the 11 ARs comprising our sample.
}             
\label{table:1}      
\centering                          
\begin{tabular}{c c c c c c c c}        
\hline                
\multicolumn{8}{c}{FLARE DURATION (data)}\\
\hline                
\multicolumn{2}{c}{Single Power Law}& \multicolumn{4}{c}{Double Power Law}& \multicolumn{2}{c}{Power Law w Exponential Rollover}\\
\hline                
  & & \multicolumn{2}{c}{Flat Power Law}   & \multicolumn{2}{c}{Steep Power Law}&  &  \\    
\hline                
 PL Index   & $Probability$ & PL Index   & $Probability$ & PL Index   & $Probability$ & PL Index   & $Probability$ \\    
\hline                        
$-1.70\pm{0.12}$	&$0.98$	&$-1.67\pm{0.09}$	&$0.98$	&$-3.37\pm{0.25}$	&$0.95$	&$-1.28\pm{0.11}$	&$0.96$	\\
\hline                                   
\end{tabular}
\end{table*}
%

%
\begin{table*}
\caption{Power law indices and respective chi-square probabilities derived by fitting several functions
to the flare durations
derived from the merged model data for the 11 ARs comprising our sample.
}             
\label{table:1}      
\centering                          
\begin{tabular}{c c c c c c c c}        
\hline                
\multicolumn{8}{c}{FLARE DURATION (merged model data)}\\
\hline                
\multicolumn{2}{c}{Single Power Law}& \multicolumn{4}{c}{Double Power Law}& \multicolumn{2}{c}{Power Law w Exponential Rollover}\\
\hline                
  & & \multicolumn{2}{c}{Flat Power Law}   & \multicolumn{2}{c}{Steep Power Law}&  & \\    
\hline                
 PL Index   & $Probability$ & PL Index   & $Probability$  &PL Index   & $Probability$ & PL Index   & $Probability$  \\    
\hline                        
$-2.79\pm{0.22}$	&$0.97$& $-1.78\pm{0.27}$	&$0.98$		&$-3.91\pm{0.42}$	&$0.94$ &$-1.13\pm{0.12}$	&$0.96$	\\
\hline                                   
\end{tabular}
\end{table*}

Although our findings show good alignment both with previous models and observations, it is crucial to crosscheck the physical soundness of our algorithm. As mentioned in Sect. 3.4, loading rule (10) does not by itself guarantee that the magnetic field remains divergence-free during the entire simulation. $WNDB$ is therefore determined in order to monitor the magnetic field divergence throughout the loading and redistribution process. Figure 11 presents the evolution of $WNDB$ during the $3\times{10}^5$ timesteps of our simulation for $AR10247$. In the beginning $WNDB$ is close to zero, as our initial condition is the extrapolated NLFF (and therefore approximately divergence-free) magnetic field. As time elapses, $\vec{\nabla\cdot\vec{B}}$ starts deviating from zero, but $WNDB$ remains under $0.20$ during the entire simulation. This holds for all ARs in our sample. Therefore, our model retains the magnetic field approximately divergence-free throughout the simulation.

Furthermore, it is worth investigating whether the use of alternative threshold definitions incurs any qualitative changes in the presented results. As an example, we apply the second threshold definition of Sect. 3.2 to $AR10247$. In this case $G_{cr}=G_{av_{max}}(1-s)\simeq{30G}$. Figure 12 shows that even with this threshold definition,
$\bar{G_{av}}$ increases gradually until timestep $1.3x10^{5}$, after which the SOC state is reached.
   $\bar{G_{av}}$ stabilizes around approximately $29.95$, which is
   lower than the critical threshold $G_{cr}=30$. The statistical properties of the generated distribution functions remain unchanged. This is also valid when switching from a $32\times{32}\times{32}$ grid towards larger volumes (, e.g. a $64\times{64}\times{64}$ grid).

   \begin{figure*}
   \centering
  \includegraphics[width=0.5\textwidth]{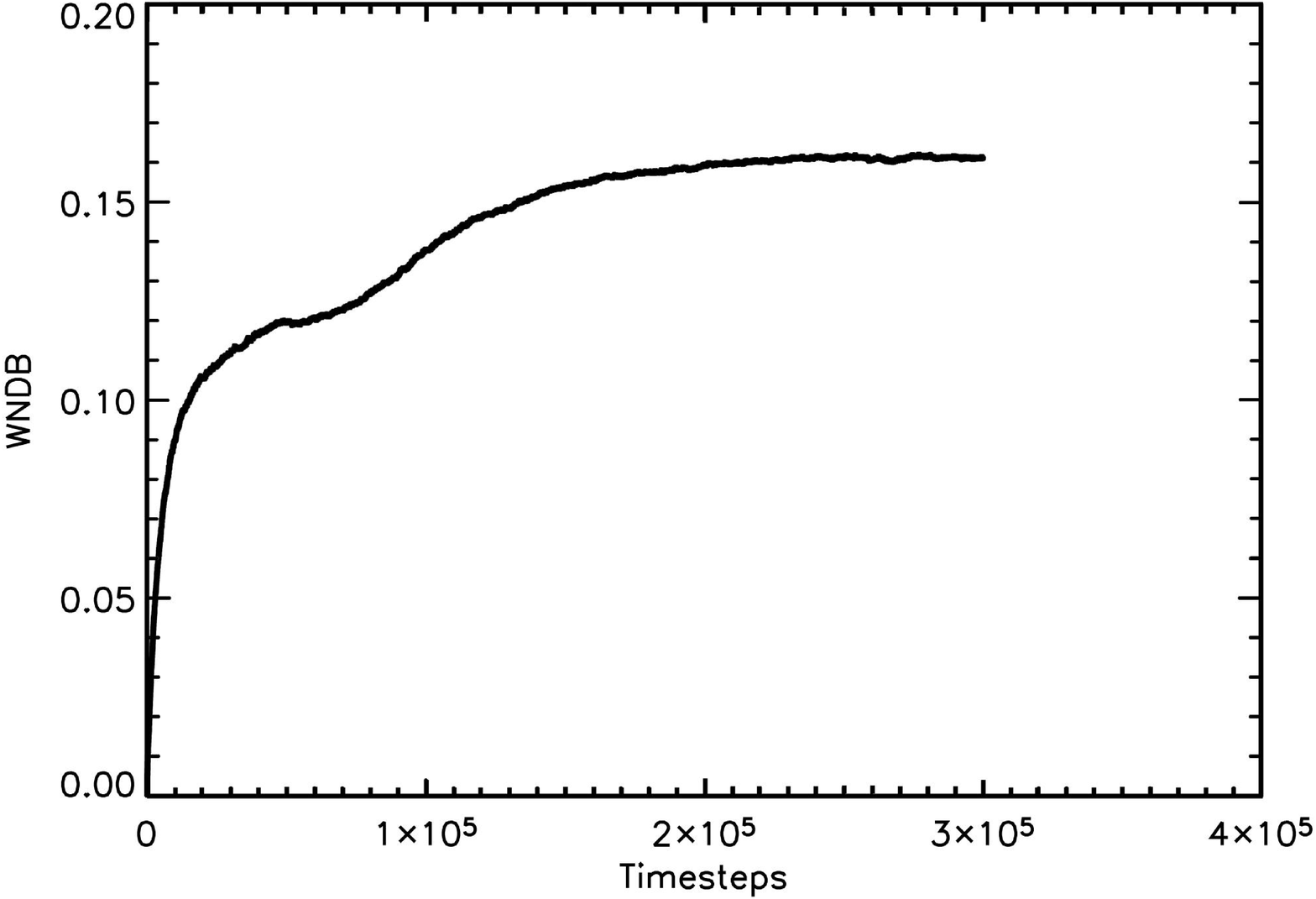}
   \caption{Evolution of $WNDB$ during the $300000$ timesteps of our simulation for $AR10247$.}
              \label{FigGam}%
    \end{figure*}

   \begin{figure*}
   \centering
  \includegraphics[width=0.5\textwidth]{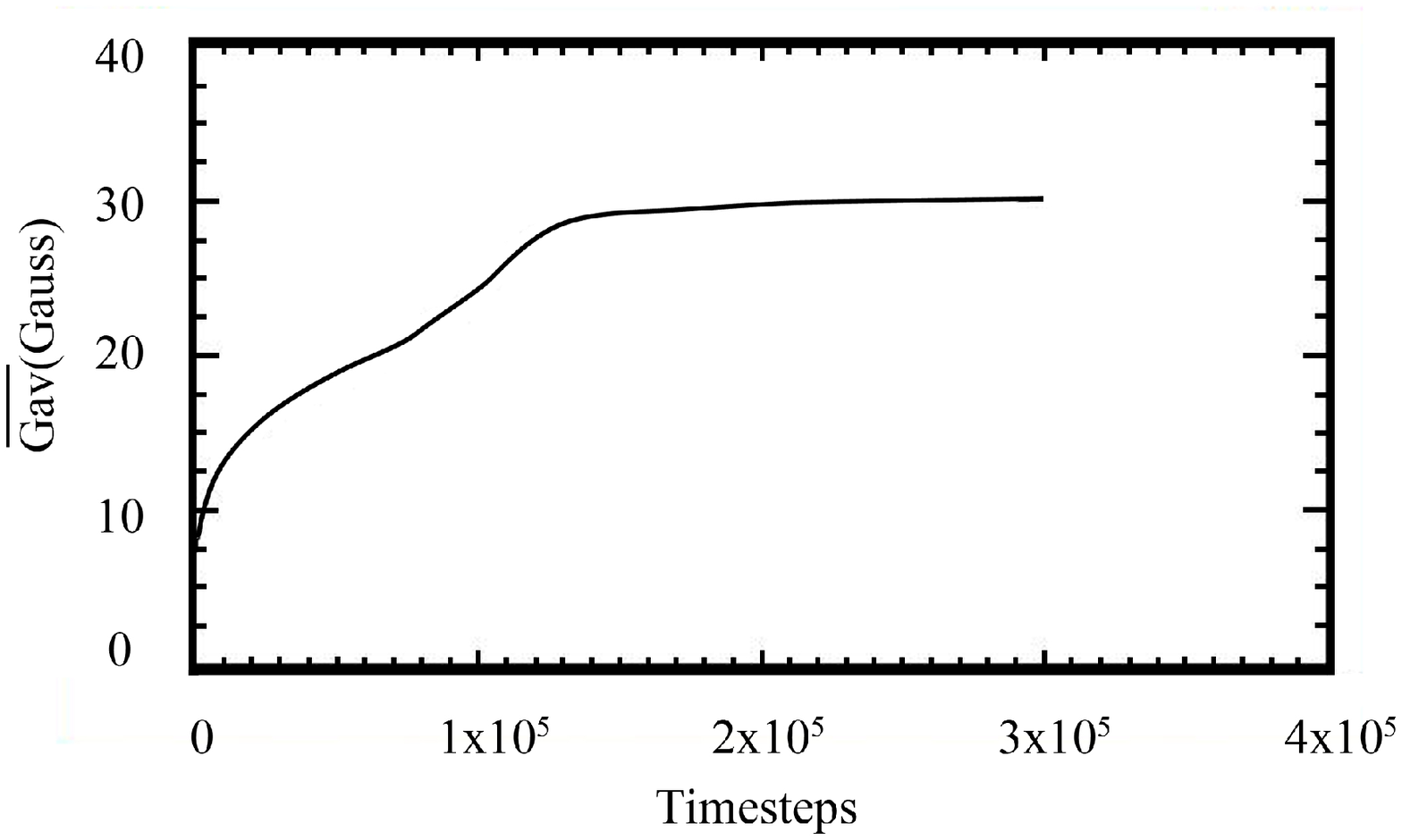}
   \caption{Diagram of the average $\bar{G_{av}}$ value over the grid for $3x10^{5}$ timesteps for $AR10247$ when the threshold definition
   is $G_{cr}=G_{av_{max}}(1-s)\simeq{30G}$.}
              \label{FigGam}%
    \end{figure*}

\section{Discussion and conclusions}

This study simulates the flaring activity of 11 solar ARs in terms of a refined CA model.
The modules comprising this integrated flare model are summarized below:

\begin{enumerate}
  \item We extrapolate the magnetic field from the photospheric boundary of 11 IVM magnetograms resampled on a $32\times{32}$ grid,
        through a nonlinear force-free optimization algorithm with preprocessing at the photospheric level (module ``EXTRA", refer to Sect. 3.1 for details).
  \item We identify the unstable locations which will dissipate magnetic energy in our grid when the approximated magnetic field Laplacian $G_{av}(\vec{r})$ at site $\vec{r}$ exceeds a specific threshold $G_{cr}$ (module ``DISCO", refer to Sect. 3.2 for details).
	 \item In case magnetic discontinuities are identified (either directly after the initialization or after each loading), the magnetic energy is redistributed such that the instabilities are completely relaxed (module "RELAX" refer to Sect. 3.3 for details).
     \item Further loading within our system is allowed, when a previously triggered avalanche has completely decayed. In this case, LOAD adds a random magnetic field increment $\vec{\delta{B(\vec{r})}}$ at a random site $\vec{r}$ within our grid according to the rules described in Sect. 3.4.
\end{enumerate}

The algorithm is allowed to run for $3x10^{5}$ timesteps, which is sufficient for all simulated
ARs to both reach the SOC state and provide sufficient event statistics after the SOC state has been reached.
The enhancements of our flare simulation model in comparison with previous SOC models of solar flares are following:

\begin{itemize}
	 \item The initial boundary conditions are not arbitrary, but stem from real solar magnetograms. An NLFF field extrapolation is used to reconstruct the initial magnetic configuration generated from the observed 11 ARs, retaining to the best possible extent physical requirements such as the minimization of the Lorentz force and the magnetic field divergence.
	 \item Given that the simulation commences from observed magnetograms, it is now possible for our CA model to remove the restriction of arbitrary energy units (see e.g. the remarks within \cite{geo01}). This gives us the opportunity to directly compare the model with the observed energy content per flare, thus leaving time as the only arbitrary quantity in our simulation.
  \item Our model follows to a significant degree the principles of \cite{luh91}. The rules obeyed during both the magnetic energy redistribution and the further driving of the system are designed in such a way that the magnetic field divergence is within tolerated limits. This has not been the case in the early CA models (Vlahos 1995, Georgoulis \& Vlahos 1996, Georgoulis \& Vlahos 1998) and has only been touched in advanced CA approaches through the use of the vector potential $\vec{A}$ instead of the magnetic field $\vec{B}$ in combination with an improved way of calculating the derivatives (Isliker et al. 2000, Isliker et al. 2001).
  \item The driving mechanism attempts to mimic not only photospheric
convection as proposed by Parker (1988, 1989, 1993), but also coronal
evolution, such as turbulence and current sheet interaction. In this
sense, locations throughout the simulation box are randomly chosen to be
perturbed. Turbulence, via
either localized Alfven waves or larger-scale turbulent flows (\cite{ein96,rap08}) leads to current-sheet interaction
that, depending on the local magnetic conditions, may trigger an
avalanche observed as a flare. Naturally, though, due to the larger
accumulation of magnetic free energy close to the photosphere (\cite{reg07})
photospheric convection and systematic photospheric motions
(e.g. shear) should be the drivers of most coronal instabilities. Our
driving mechanism should be further revised to account for systematic
photospheric flows. This future step is important because (1) it has
been argued that the distribution and energy content of magnetic
discontinuities in a given photospheric boundary can explain the
statistical properties of flares (\cite{vla04}) and (2)
investigating possible correlations between the photospheric driver and the corresponding coronal active region reveals the strong nonlinearity of active-region magnetic
configurations that hinders correlations between photospheric and coronal structures (Dimitropoulou et al. 2009). The latter patterns,
however, have a crucial impact on the expected dynamical activity of the
system, namely, the magnetic energy release and the subsequent particle
acceleration processes (\cite{vla04b}).
  \item The derived results can be directly compared with flare observations, due to the fact that the simulation uses extrapolated fields from observed vector magnetograms as initial conditions.
       At this point, we once again stress that the X-ray flares recorded by GOES for each AR do not comprise a statistically reliable sample. Therefore, in order to make such a comparison possible, we merged the GOES flare data of all 11 ARs in our sample into one database, comprising of $154$ flares.
\end{itemize}

Our results show that under the imposed driving and redistribution rules, all examined ARs reach the SOC state. The retrieved distribution functions for event duration are best described by either double power laws or power laws with
exponential rollover, although single power laws are also applicable for the merged data. The peak energy and total energy follow clearly
single power laws. The power law indices for durations and energies as presented in Tables 1,2 and 3 lie in the well-known ranges documented
consistently in numerous past studies, including \cite{geo01}. In this study, Georgoulis et al. compare their SOC model with data from the
Danish Wide Angle Telescope for Cosmic Hard X-rays (WATCH) collected during maximum of the solar cycle 21. Figure 1 in the cited work shows
that the peak and total energy of the observed flares follow single power law distribution functions with indices $-1.59$ and $-1.39$
correspondingly, whereas the flare duration distribution function is considered to either follow double power law (with index $-1.15$
for the flat and $-2.25$ for the steep part) or power law with exponential rollover (with power law index $-1.09$). These results are in
agreement with our findings. Although our model generates flare duration distribution functions with indices in alignment with the ones
presented in \cite{geo98}, we did not attempt here to reproduce two key findings of \cite{geo98}, namely the variability of the scaling
indices as a result of the  driver's variability as well as the two distinct event populations. In the cited study the peak and total
energy distribution functions follow double power laws. The steeper part of them corresponds to the signature of a ``soft" flare population
(nanoflares), whereas the flatter part is attributed to microflares and flares.

Although this work overcomes major drawbacks of many previous CA models, such as retaining the value of the magnetic field divergence close to zero throughout the simulation, there are still some points that can lead to discrepancies. First and foremost, the determination of the threshold value $G_{cr}$ can slightly influence the exponents of the retrieved power laws, although it cannot cause any qualitative change to their appearance, namely the known flare statistical properties will always follow power law distributions, independent of the threshold value imposed. The histogram method presented in Sect. 3.2 eliminates to an extent the arbitrary selection of $G_{cr}$. We have also investigated whether the rebinning of our grid to the size of $32\times{32}\times{32}$ influences our results in comparison with larger grids (e.g. $64\times{64}\times{64}$) and we found that the differences in the power law indices lie within the inferred uncertainties. Finally, as far as the comparison with the observational data is concerned, we have already stressed that this is a preliminary attempt given that the number of GOES X-ray flares across all investigated ARs does not produce sufficient statistics.

The discussion regarding the validity of the CA models when it comes to the simulation of physical processes in complex systems is a long-running one. As discussed by \cite{isl98}, the essence of CA modeling is to describe complex systems, which comprise a large number of interacting subsystems, assuming that the global dynamics described statistically are not sensitive to the fine structure of the elementary processes. More strict approaches such as MHD, on the other hand, are based on the precise description of the elementary processes through detailed differential equations. Both approaches have been shown to exhibit drawbacks and advantages. The CA approach does not provide any insight into the local processes or over short time intervals, but it reproduces the global statistics. MHD reveals details about the local processes, but coupling them to a global description is a formidable task. In this sense, the two approaches are complementary and there have been indeed various attempts to either combine them (e.g. \cite{lon00}), or interpret CA models as discretized MHD equations (Isliker et al. 1998,Vassiliadis et al. 1998). Even more extended CA models, like the X-CA model described by Isliker et al. (2001), have achieved consistency with MHD to a greater extent. Our CA model will opt to incorporate and utilize meaningful modeling developments into a more concrete, ``integrated" flare model.

\begin{acknowledgements}
      We are grateful to Thomas Wiegelmann who kindly contributed his NLFF extrapolation code to this analysis. Xenophon Moussas and Dafni Strintzi are acknowledged
      for their constructive comments in the course of this work. Additionally, IVM Survey data are made possible through the staff of the U. of Hawaii, supported by the Air Force Office of Scientific Research, contract F49620-03-C-0019. IVM Survey data
are based upon work supported by the National Science Foundation under Grant No. 0454610. Any opinions, findings, and conclusions or recommendations expressed in this
material are those of the author(s) and do not necessarily reflect the views of the National Science Foundation (NSF). Finally, we thank the anonymous referee whose help significantly improved the paper.
\end{acknowledgements}


\begin{thebibliography}{}


\bibitem[Bak et al. 1987]{bak87} Bak, P., Tang, C., \& Wiesenfeld, K. 1987, \prl, 59, 381
\bibitem[Biesecker 1994]{bie94} Biesecker, D. A. 1994, Ph. D. Thesis, University of New Hampshire
\bibitem[Bromund et al. 1995]{bro95} Bromund, K. R., McTiernan, J. M., \& Kane, S. R. 1995, \apj, 455, 733
\bibitem[Crosby et al. 1993]{cro93} Crosby, N. B., Aschwanden, M. J., \& Dennis B. R. 1993, \solphys, 143, 275
\bibitem[Datlowe et al. 1974]{dat74} Datlowe, D. W., Elcan, M. J., \& Hudson, H. S. 1974, \solphys, 39, 155
\bibitem[Dennis 1985]{den85} Dennis, B. R. 1985, \solphys, 100, 465
\bibitem[DeRosa et al. 2009]{sch09} DeRosa, M. L., Schrijver, C. J., Barnes, G., Leka, K. D., Lites, B. W., Aschwanden, M. J., Amari, T., Canou, A., McTiernan, J. M., Regnier, S., Thalmann, J. K., Valori, G., Wheatland, M. S., Wiegelmann, T., Cheung, M. C., Conlon, P. A., Fuhrmann, M., Inhester, B., \& Tadesse, T. 2009, \apj, 696, 1780
\bibitem[Dimitropoulou et al. (2009)]{dim09} Dimitropoulou, M., Georgoulis, M., Isliker, H., Vlahos, L., Anastasiadis, A., Strintzi, D., \& Moussas, X. 2009, \aap, 505, 1245
\bibitem[Einaudi et al. 1996]{ein96} Einaudi, G., Velli, M., Politano, H., \& Pouquet, A. 1996, \apj, 457, L113
\bibitem[Galsgaard (1996)]{gal96} Galsgaard, K. 1996, \aap, 315, 312
\bibitem[Georgoulis et al. (1995)]{geo95} Georgoulis, M. K., Kluiving, R., \& Vlahos, L. 1995, Physica A, 218, 191
\bibitem[Georgoulis \& Vlahos (1996)]{geo96} Georgoulis, M. K., \& Vlahos, L. 1996, \apj, 469, L135
\bibitem[Georgoulis \& Vlahos (1998)]{geo98} Georgoulis, M. K., \& Vlahos, L. 1998, \apj, 336, 721
\bibitem[Georgoulis et al. (2001)]{geo01} Georgoulis, M. K., Vilmer, N. \& Corsby, N. B. 2001, \aap, 367, 326
\bibitem[2005]{geo05} Georgoulis, M. K. 2005, \apj, 629, 69
\bibitem[Harvey \& Zwaan 1993]{har93} Harvey, K. L., \&  Zwaan, C. 1993, \solphys, 148, 85
\bibitem[Isliker et al. (1998)]{isl98} Isliker, H., Anastasiadis, A., Vassiliadis, D.,  \& Vlahos, L. 1998, \aap, 335, 1085
\bibitem[2000]{isl00} Isliker, H., Anastasiadis, A., \& Vlahos, L. 2000, \aap, 363, 1134
\bibitem[2001]{isl01} Isliker, H., Anastasiadis, A., \& Vlahos, L. 2001, \aap, 377, 1068
\bibitem[(2002)]{isl02} Isliker, H., Anastasiadis, A., \& Vlahos, L. 2002, ESASP, 506, 6411
\bibitem[Lin et al. 1984]{lin84} Lin, R. P., Schwartz, R. A., Kane, S. R., Pelling, R. M., \& Hurly, K. C. 1984, \apj, 283, 421
\bibitem[Longope \& Noonan 2000]{lon00} Longope, D. W., \& Noonan, E. J. 2000, \apj, 542, 1088
\bibitem[Lu \& Hamilton (1991)]{luh91} Lu, E. T.,\& Hamilton, R. J. 1991, \apj, 380, L89.
\bibitem[Lu et al. (1993)]{luh93} Lu, E. T., Hamilton, R. J., McTiernan, J. M., \& Bromund, K. R. 1993, \apj, 412, 841
\bibitem[MacKinnon \& Macpherson (1997)]{mac97} MacKinnon, A. L., \& Macpherson, K. P. 1997, \aap, 326, 1228
\bibitem[Mickey et al. 1996]{mic96} Mickey, D. L., Canfield, R. C., LaBonte, B. J., Leka, K. D., Waterson, M. F., \& Weber, H. M. 1996,
\solphys, 168, 229M
\bibitem[1988]{par88} Parker, E. N. 1988, \apj, 330, 474
\bibitem[1989]{par89} Parker, E. N. 1989, \solphys, 121, 271
\bibitem[1993]{par93} Parker, E. N. 1993, \apj, 414, 389
\bibitem[Polygiannakis et al. 2002]{pol02} Polygiannakis, J. M., Nikolopoulou, A., Preka-Papadima, P., Moussas, X., \& Hilaris, A. 2002, ESASP, 505, 541
\bibitem[Priest et al. 2003]{pri03} Priest, E. R., Hornig, G., \& Pontin, D. I. 2003, J. Geophys. Res. Space Phys.,
108(A7), 1285
\bibitem[Rappazzo et al. 2008]{rap08} Rappazzo, A. F., Velli, M., Einaudi, G., \& Dahlburg, R. B. 2008, \apj, 677, 1348
\bibitem[Regnier \& Priest 2007]{reg07} Regnier, S., \&  Priest, E. R. 2007, \aap, 468, 701
\bibitem[Shimizu 1995]{shi95} Shimizu, T. 1995, PASJ, 47, 251
\bibitem[Sturrock et al. 1984]{stu84} Sturrock, S., Kaufmann, P., Moore, P. L., \& Smith, D. F. 1984, \solphys, 94, 341
\bibitem[Vilmer 1987]{vil87} Vilmer, N. 1987, \solphys, 111, 207
\bibitem[Vlahos et al. (1995)]{vla95} Vlahos, L., Georgoulis M. K., Kluiving R., \& Paschos P. 1995, \aap, 299, 897
\bibitem[Vlahos \& Georgoulis 2004]{vla04} Vlahos, L., \&  Georgoulis, M. K. 2004, \apjl, 603, 61
\bibitem[Vlahos et al. 2002]{vla02} Vlahos, L., Fragos, T., Isliker H., \& Georgoulis M. K. 2002, \apjl, 575, 87
\bibitem[Vlahos et al. 2004]{vla04b} Vlahos, L., Isliker H., \& Lepreti, F. K. 2004, \apj 608, 540
\bibitem[Vassiliadis et al. 1998]{vas98} Vassiliadis, D., Vlahos, L., \& Georgoulis, M. 1998, \apj, 509, L53
\bibitem[Wheatland et al. 2000]{whe00} Wheatland, M. S., Staurrock, P. A., \& Roumeliotis, G. 2000, \apj, 540, 1150
\bibitem[Wiegelmann 2004]{wie04} Wiegelmann, T. 2004, \solphys, 219, 87
\bibitem[Wiegelmann et al. 2006]{wie06} Wiegelmann, T., Inhester, B., \& Sakurai, T. 2006, \solphys, 233, 215
\bibitem[Wiegelmann 2008]{wie08} Wiegelmann, T. 2008, \jgr, 113, A03S02
\end{thebibliography}
\end{document}